\documentclass[journal, amsmath, amsfonts]{IEEEtran}
\usepackage{graphicx}
\usepackage{float}
\usepackage{threeparttable}
\usepackage{subfigure}
\usepackage{amssymb}
\begin{document}
\title{Heat transfer and SET voltage in filamentary RRAM devices}

\author{Dipesh~Niraula, and~Victor~G.~Karpov
\thanks{Manuscript received March 3, 2017. This work was supported in part by the Semiconductor Research Corporation (SRC) under Contract No. 2016LM-2654.}
\thanks{D.~Niraula and V.~G.~Karpov are with the Department of Physics and Astronomy, University of Toledo, Toledo, OH, 43606 USA (e-mail: dipesh.niraula@rockets.utoledo.edu; victor.karpov@utoledo.edu).}}
\maketitle

\begin{abstract}
We study the heat transport in filamentary RRAM nano-sized devices by comparing the accurate results of COMSOL modeling with simplified analytical models for two complementary mechanisms: one neglecting the radial heat transfer from the filament to the insulating host, while the other describing the radial transport through the dielectric in the absence of the filament heat transfer. For the former, we find that the earlier assumed simplification of the electrodes being ideal heat conductors is insufficient; a more adequate approximation is derived where the heat transport  is determined by the adjacent proximities of the filament tips in the electrodes.  We find that both complementary mechanisms overestimate the maximum temperature yet offering acceptable results. However, the two in parallel provide a better analytical approximation. In addition, we show that the Wiedemann-Franz-Lorenz law helps the analysis when the Lorenz parameter is chosen from the actual data.  We present an approximate expression for the SET voltage possessing a high degree of universality and predicting that filament materials with low Lorenz numbers can be good candidates for the future low set voltage devices.
\end{abstract}


\section{Introduction}
Modern resistive random access memory (RRAM) devices are about 100-300 nm thick including two metal electrodes and the the insulator layer of about 10-70nm thickness. For the the insulator thickness of $\sim 10$ nm and and bias voltage of $\sim 1$ V, the electric field strength in the insulator $E \sim 10^8$ V/m. In response to that field, a conductive path through the insulator can be created in the form of a conductive filament. The regime of continuous conductive filament corresponds to the device ON state created at the SET transition and discontinued at the RESET transition.  When the current flows through the filament, a significant Joule heat is liberated creating a temperature distribution that affects the device operations. Here, we present an analysis of that temperature distribution with the emphasis on its underlying physics facilitating the device understanding and design. This is achieved by (1) numerically generating the temperature distribution by means of COMSOL multiphysics package and (2) comparing the latter with two analytically solved limiting cases: (a) heat transport to the electrodes dominated by the metal filament without lateral spreading , vs. (b) that dominated by the lateral spreading of thermal energy preceding its dissipation at the electrodes. Based on our conclusions, we present a related expression for the filamentary RRAM SET voltage that possesses a high degree of universality.
\begin{figure}[t]
  \centering
  \includegraphics[width=0.45\textwidth]{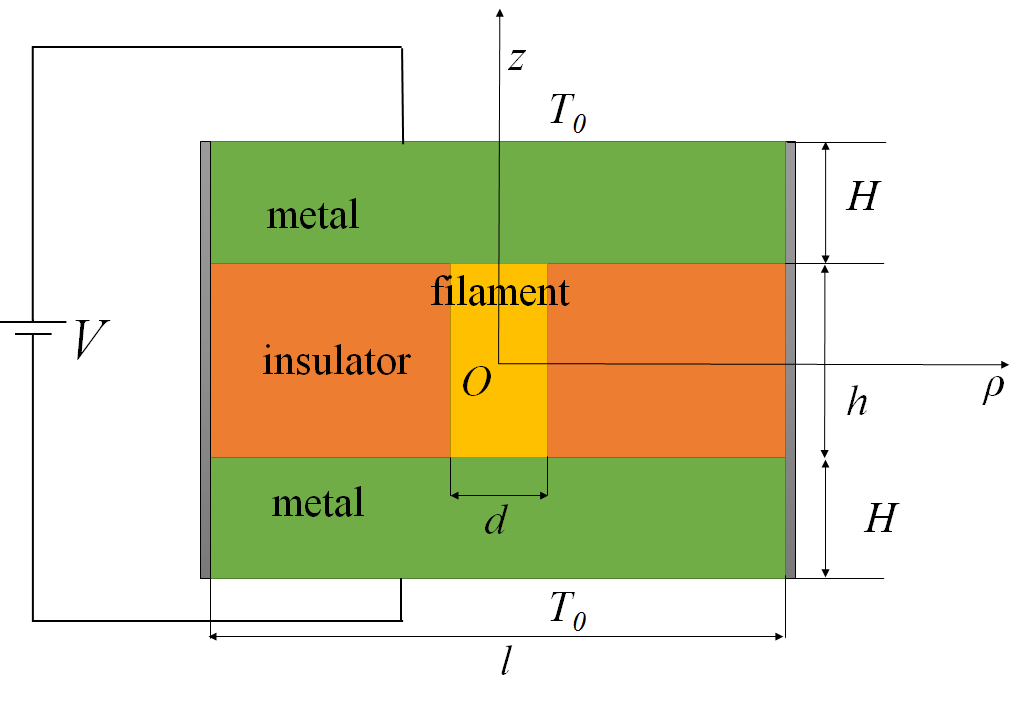}
  \caption{Parameters describing the geometry of a simple cylindrical RRAM device which consist of two metal electrodes, insulator layer, and a metallic filament. The device is thermally insulated except for the top and bottom boundary of the metal layer which is fixed at an ambient temperature $T_0$ and is subjected to an external bias of $V$ volts.  We fix the origin of our cylindrical co-ordinate system at the center of the filament. Additional material parameters not shown in the figure are the thermal and electrical conductivities of the metal, filament, and insulator layer which are respectively ($\kappa_m$, $\sigma_m$), ($\kappa_f$, $\sigma_f$), and ($\kappa_i$, $\sigma_i$) required to solve the heat transfer problem. Note: Figure not drawn to scale($H>h$)}\label{Fig:MIM}
\end{figure}

\section{Numerical Modeling}\label{sec:num}
We use an axillary symmetric model of a RRAM device, the cross-section of which is  shown in Fig.\ref{Fig:MIM}; a description of its COMSOL modeling routine is presented in Appendix \ref{appendix0}. The dielectric layer corresponds to the metal oxide HfO$_2$. (As widely accepted, the oxygen sweeps creating oxygen vacancies that form a conductive filament of a non-stoichiometric composition HfO$_{2-x}$.) Following a number of implementations \cite{Govoreanu 2013,Fantini 2012}, we choose TiN for the metal electrode materials. The material properties are listed in Table \ref{tab:param}.

\begin{table}[t]\footnotesize
\centering
\caption{Material parameters  used in numerical modeling\cite{Panzer 2009,Hildebrandt 2011,Govoreanu 2013,Samani 2013,Yaw 2015,Lide 2008}}\label{tab:param}
    \begin{threeparttable}
            \begin{tabular}{ |c | c | c | c | c |}
                 \hline
                    Material                                 & TiN      & HfO$_2$ & HfO$_{2-x}$    & Hf \\ \hline
                    Thermal Conductivity[W/K.m]              & 11.9     & 0.5     & 20\tnote{a}    & 23 \\
                    Electrical Conductivity[S/m]             & 10$^6$   & 10$^{-2}$ & 10$^5$       & 3$\cdot10^6$ \\
                    Specific Heat\tnote{c} [J/kg.K]          & 545.33   & 120     & 130\tnote{a}    & 144 \\
                    Relative Permittivity                    &-$\infty$\tnote{b} & 25      &-$\infty$\tnote{a}\tnote{ ,b} &-$\infty$\tnote{b} \\
                    Density[kg/m$^3$]                        & 5220     & 9680    & 12000\tnote{a}   & 13310 \\
                    \hline
            \end{tabular}
        \begin{tablenotes}
         \item[a]Assumed value such that it lies in between Hf and HfO$_2$
         \item[b]-10$^6$ was used instead of -$\infty$ for practical purpose
         \item[c]Specific heat capacity at constant pressure
        \end{tablenotes}
    \end{threeparttable}
\end{table}

\begin{figure}
\centering
\subfigure[Longitudinal temperature distribution for device I .]
{\includegraphics[width=0.22\textwidth]{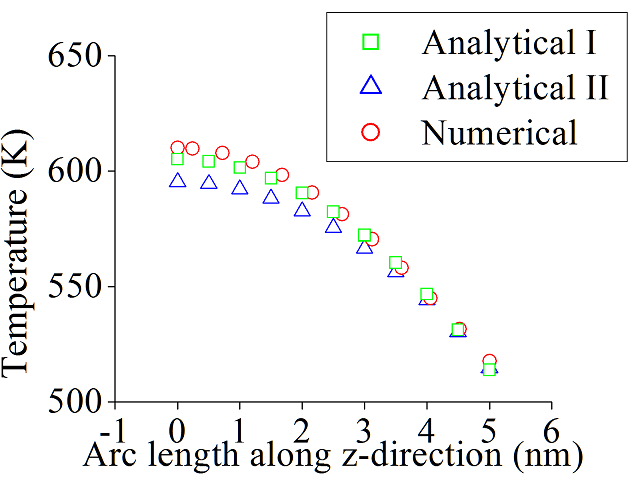}\label{Fig:COMSOLtempdistA10nm}}
\hspace{0.001\textwidth}
\subfigure[Longitudinal temperature distribution for device II]
{\includegraphics[width=0.22\textwidth]{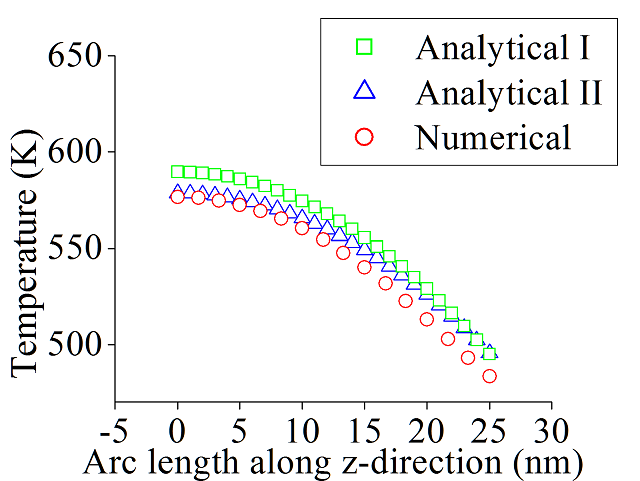}\label{Fig:COMSOLtempdistA50nm}}

\subfigure[Radial temperature distribution for device I.]
{\includegraphics[width=0.22\textwidth]{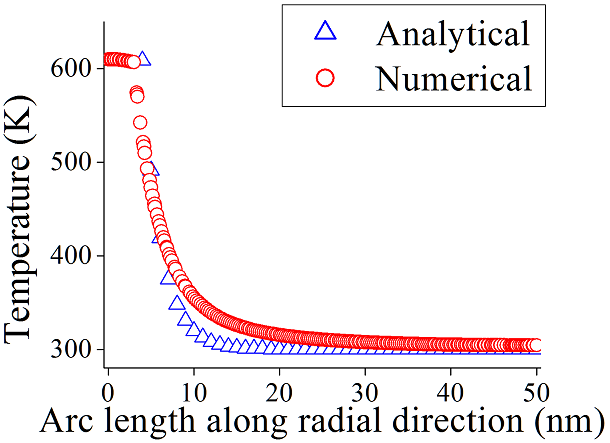}\label{Fig:COMSOLtempdistB10nm}}
\hspace{0.001\textwidth}
\subfigure[Radial temperature distribution for device II.]
{\includegraphics[width=0.22\textwidth]{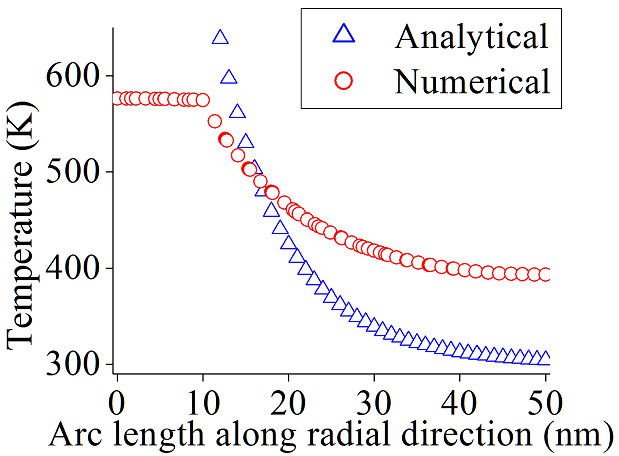}\label{Fig:COMSOLtempdistB50nm}}
    \caption{Temperature distribution along the Longitudinal and Transverse direction in the mid-layer of the metal-insulator-metal structured RRAM device. In the longitudinal temperature distribution plots, Figs. \ref{Fig:COMSOLtempdistA10nm} and \ref{Fig:COMSOLtempdistA50nm}, Analytical I curve corresponds to Eq.(\ref{eq:Parabolic}) and Analytical II curve corresponds to Eq.(\ref{eq:TempDist1}) while in the radial temperature distribution plots, Figs. \ref{Fig:COMSOLtempdistB10nm} and \ref{Fig:COMSOLtempdistB50nm}, Analytical curve corresponds to Eq.(\ref{eq:tempdist})}.
    \label{Fig:COMSOLtempdist}
\end{figure}

\begin{table}[t]\footnotesize
\centering
\caption{Device dimensions used in numerical modeling}\label{tab:dimen}
    \begin{threeparttable}
            \begin{tabular}{ |c | c | c | c | c |}
                 \hline
                    Device/Dimension (nm)                                  & $H$     & $h$ & $l$   & $d$ \\ \hline
                    Device I              & 30     & 10     & 100   & 6 \\
                    Device II             & 100   & 50 & 100       & 20 \\

                    \hline
            \end{tabular}
            \end{threeparttable}
\end{table}
Two device dimensions chosen for modeling are described in Table \ref{tab:dimen}. The top electrode is connected to a source voltage of 0.5 V while the bottom one is grounded. An ambient temperature of 300K is maintained at the top and bottom. The side walls of the device are taken to be electrically and thermally insulated, which reflects an array geometry with zero inter-device currents. With those boundary conditions, COMSOL solves the coupled heat and electrostatic equations producing the temperature distributions presented in Fig. \ref{Fig:COMSOLtempdist}.

As shown in Figs. \ref{Fig:COMSOLtempdistA10nm}, \ref{Fig:COMSOLtempdistA50nm}, \ref{Fig:COMSOLtempdistB10nm}, and \ref{Fig:COMSOLtempdistB50nm}, the temperature distributions are a maximum of about 610 K and 576 K at the center of the filament for the Devices I and II respectively. The temperature decay scales in the axial and lateral directions are comparable, although the functional forms are different. The similarity of the two decay scales is due to the mutually balancing competing factors: a better thermal conductivity in a metal vs. the greater area facing the dielectric. In conceivable cases of different material parameters or geometrical dimensions, one of those factors can dominate.

\section{Limiting cases}\label{sec:lim}
For comparison, we present below the two limiting cases corresponding to (a) the 1D heat transfer along the filament and (b) the dominating radial transport of heat from the filament to the insulator. At the end, we combine the two limiting cases as the thermal resistance in parallel to obtain a single heat transport model.

\subsection{1D Thermal Transport}\label{sec:1D}
Assuming the thermal conductivity of the insulating layer much less than that of the filament material, the heat will flow only through the filament corresponding to a 1D problem. Consider a filament of length $h$ embedded in an insulator sandwiched between two metal electrodes and subjected to an external voltage V as shown in Fig. (\ref{Fig:MIM}). The ends of the filament are at the filament-junction temperature $T_j$ which depends on the ambient temperature $T_0$ maintained at the surface of the metal electrodes.

Consider a steady state heat equation with Joule heat as a source term,
\begin{equation}\label{eq:heateq1}
-\nabla(\kappa(T)\nabla T)= \bf{J}\cdot\bf{E}.
\end{equation}
where, $\bf{J}$ is the current density and $\bf{E}$ is the electric field in the filament. Since $\delta\kappa/\kappa \ll \delta T/T$, we treat thermal conductivity to be a constant; a broader analysis is presented in \cite{John1985}. Using the Ohms law, $\bf{J}= \sigma\bf{E}$, the above equation takes the form,
\begin{equation}\label{eq:heateq2}
\nabla^{2}T = -\frac{\sigma}{\kappa}E^{2}\equiv -\frac{E^{2}}{L}\frac{1}{T}.
\end{equation}
Here we have accounted for the  Wiedemann-Franz-Lorenz Law\cite{W-F 1853} ,\cite{Lorenz 1872}, stating that  $\sigma/ \kappa=1/LT$ where $L$ is the Lorenz number. Sommerfeld\cite{sommerfeld 1927}, showed that for `good' metals, $L = (\pi k_B)^2/3e^2 = 2.44 \cdot 10^{-8}$ W$\Omega$K$^{-2}$ where, $k_B$ is the Boltzmann's constant and $e$ is the electron charge.

In reality, the measured $L$ varies between different metals metals and can be temperature dependent, off from the Sommerfeld's value within an order of magnitude. \cite{kumar 1993} In particular, for the case under consideration, the values from Table \ref{tab:param} lead to $L\approx 6.67\cdot 10^{-7}$ W$\Omega$K$^{-2}$, significantly different from the Sommerfeld's prediction. That may indicate that the filament material is not a homogeneous metal representing perhaps a mixture of metallic and insulating phases \cite{kruchihin2015}. Earlier, the Wiedemann-Franz-Lorenz relation was used in modeling of phase change memory devices \cite{kencke2007}.

Eq. (\ref{eq:heateq2}) has a simple solution for the regime of low heat, $\delta T\ll T$, predicting the parabolic temperature distribution,
\begin{equation}\label{eq:Parabolic}
T(z) = T_j + \frac{E^2h^2}{8LT_j}\left[1-\left(\frac{2z}{h}\right)^2\right]
\end{equation}
with the maximum (center) temperature by,
\begin{equation}\label{eq:weaklim}
\delta T=\frac{E^2h^2}{8LT_j}
\end{equation}
above the junction (at $z=\pm h/2$) temperature $T_j$.

\begin{figure}[t]
  \centering
  \includegraphics[width=0.45\textwidth]{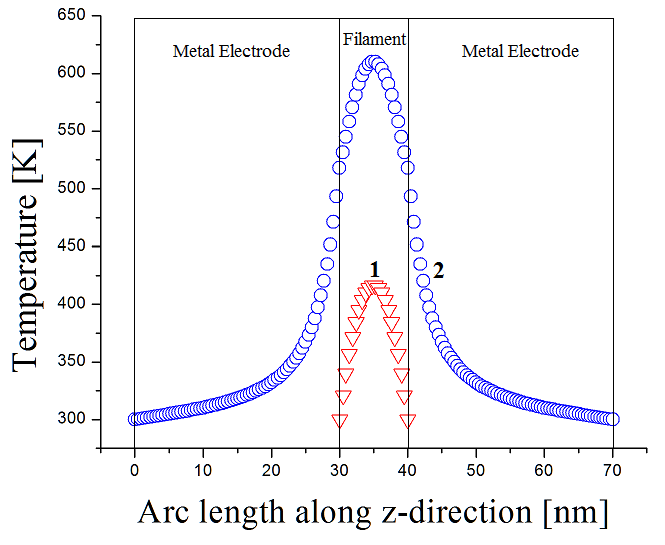}
  \caption{The COMSOL calculated temperature distributions in Device I along the z-axis. Curve 1 represents the  distribution when the room temperature (300K) is assumed at the electrode-dielectric interface; curve 2 corresponds to the case when the free surface of the electrode is at room temperature, while the electrode-dielectric interface temperature is derived as explained in the text.}\label{Fig:junctiontemp}
\end{figure}
The latter solution was earlier presented in \cite{John1985} and \cite{ielimini2011}. It was assumed \cite{ielimini2011}, that the junction filament temperature is the same as the ambient temperature,  $T_j = T_0$, which may seem intuitively justified because of the high thermal conductivity of the metal electrode . An important correction here is that $T_j$ can be significantly higher than $T_0$  because the thermal transport in the electrode is dominated by the geometrically small region adjacent to the filament tip that plays the role of the point heat source at the electrode interface. Indeed, the thermal resistance of the series of elemental co-centric semi-spherical layers centered at the tip, is determined by the small distance contributions, $r\sim d$. The significant difference between $T_j$ and $T_0$ is illustrated in Fig. \ref{Fig:junctiontemp}.

Analytically, the temperature distribution at distance $r$ in the electrode from the above point source satisfying the condition that $T(r)=T_0$ at $r\rightarrow \infty$, takes the form,
\begin{equation}\label{eq:Tr}
T_r =(T_j - T_0)\frac{d}{2r} +T_0
\end{equation}
The corresponding heat flow through a hemispherical surface of radius $d/2$ in the electrode, $2\pi (d/2)^2 \kappa_m \nabla T$ with $T$ from Eq. (\ref{eq:Tr}) must be equal to that supplied by the filament, $(2\delta T/h)(\pi d/2^2)$ as follows from Eq. (\ref{eq:Parabolic}). That equality yields,
\begin{equation}\label{eq:tempjunction}
T_j = T_0 + \frac{1}{2}\sqrt{T_0 ^2 + \frac{\kappa_f}{\kappa_m} \frac{E^2 h d}{4L}}
\end{equation}
The maximum temperature is now given by,
\begin{equation}\label{eq:Tmax}T_{\max}=T_j+\delta T\end{equation}
with $T_j$ from Eq. (\ref{eq:tempjunction}) and $\delta T$ from Eq. (\ref{eq:weaklim}).

For numerical estimate, we use  $L\approx 6.67\cdot 10^{-7}$ W$\Omega$K$^{-2}$, $T_0=300$ K, and $V=0.5$V, yielding $\delta T\equiv T_\textrm{max}-T_j\approx 91$ K and $94$ K respectively for Device I and Device II, while the junction temperatures are $T_j \approx 514$ K and $495$ K respectively, which is close to the numerical solution presented in Figs. \ref{Fig:COMSOLtempdistA10nm} and \ref{Fig:COMSOLtempdistA50nm}.

As shown in Appendix \ref{appendix1} and the earlier general treatment \cite{John1985}, Eq. (\ref{eq:heateq2}) can be integrated more accurately to result in
\begin{equation}\label{eq:TempDist1}
T(z)=T_\mathrm{max}\exp\left[-\left(\mathrm{erf}^{-1}\left(\frac{\mp z}{T_\mathrm{max}}\sqrt{\frac{2E^{2}}{\pi L}}\right)\right)^2\right]
\end{equation}
where erf$^{-1}$ is the inverse error function, and the maximum temperature is defined by the transcendental equation
\begin{equation}\label{eq:T_max}
\frac{T_\mathrm{max}}{T_j}\mathrm{erf}\left(\sqrt{\ln\left(\frac{T_\mathrm{max}}{T_j}\right)}\right) = \mp\frac{h}{2 T_j}\sqrt{\frac{2E^{2}}{\pi L}}.
\end{equation}
These results are plotted in Figs. \ref{Fig:COMSOLtempdistA10nm} and \ref{Fig:COMSOLtempdistA50nm} predicting $\delta T\equiv T_\textrm{max}-T_j\approx 85$ K and 80 K, for Device I and Device II respectively, rather close to the above COMSOL result.

\subsection{Radial temperature distribution}\label{sec:rad}

In spite of the relatively small thermal conductivity of an insulator, the heat flowing from the filament to the metal electrodes through the insulator  can be significant because of the large areas of the filament/dielectric and dielectric/electrode interfaces. To understand the corresponding radial distribution of temperature, we adopt a simplified model of a very thin insulating layer where the transversal temperature variation can be neglected, thus approximating $T$ with its average value $\overline{T}(\rho)$.

Equating the radial component of the divergence of heat flow $\kappa _ih\rho ^{-1}(\partial /\partial \rho)\rho (\partial /\partial \rho)\overline{T}$ to the heat $\kappa _m(\overline{T}-T_0)/H$ absorbed by the electrodes leads to the equation,
\begin{equation}\label{eq:divT}
\frac{1}{\rho}\frac{d}{d \rho}\rho\frac{d \overline{T}}{d \rho} + \beta ^2 (T_0 -\overline{T})=0\ {\rm with}\ \beta =\sqrt{2\frac{\kappa_m}{\kappa_iHh}}
\end{equation}
where $\beta$ plays the role of the temperature reciprocal decay length.

The solution to Eq. (\ref{eq:divT}) is expressed through the Bessel function with the coefficient determined by the condition that the heat flux $\pi hd(d\overline{T}/d\rho )$ at $\rho =d/2$ equals the filament generated Joule heat. As shown in Appendix (\ref{appendix2}), this results in
\begin{equation}\label{eq:tempdist}
\overline{T}(\rho) = \frac{V^2d}{4 h \kappa_i \beta(h/\sigma_f+2H/\sigma_m)K_1(\beta d/2)} K_0(\beta\rho) + T_0
\end{equation}
where $K_0$ and $K_1$ are the modified Bessel functions of order 1 and 0.

For numerical estimate, we use the above presented device parameters, which yields the temperature decay length $1/\beta = 2.51$ nm and 10.24 nm for Device I and Device II respectively.
The resulted temperature distribution inside the insulator is plotted in Fig. \ref{Fig:COMSOLtempdistB10nm} along with the numerically solved (COMSOL) temperature distribution. A somewhat overestimated temperature of the filament is due to the approximation neglecting thermal exchange through the filament bases.


\subsection{Parallel Transport}
A more accurate temperature estimate can be obtained by considering the parallel connection of the thermal resistance from 1D heat transport $R_{th1D}$, and that of radial heat transfer, $R_{thrad}$, which are derived from respectively Eq.(\ref{eq:weaklim}) and Eq.(\ref{eq:tempdist}),
\begin{eqnarray}\label{eq:thermalRes}
R_\mathrm{th1D}&=&\frac{R}{8LT_j}\quad \mathrm{and}\\
R_\mathrm{thrad} &=&\frac{R d K_0(\beta d/2)}{4 h \kappa_i \beta(h/\sigma_f+2H/\sigma_m)K_1(\beta d/2)}
\end{eqnarray}
where $R$ is the electrical resistance. The parallel connection results in a smaller total thermal resistance, $ 1/R_\textrm{th} = 1/ R_\textrm{th1D} + 1/ R_\textrm{thrad}$.

The maximum temperature can be calculated using $\delta T = R_\textrm{th} \cdot V^2/R$. For numerical estimate of Device I and Device II, we use the above presented  parameters, which yields $\delta T \equiv T_\textrm{max}-T_j\approx 82$ K and 79 K, respectively. In Sec. \ref{sec:lim}, we obtained $T_j$ to be 514 K and 495 K for Device I and II respectively, using which the maximum temperature, $T_\textrm{max}$, of the filament is found to be 596 K and 574 K which is yet closer to the COMSOL numerical solution.


\section{SET Voltage}
Our thermodynamic analysis of resistive switching relates $V_{\textrm{SET}}$ as in [\cite{Karpov2016}, eq.(8)],
\begin{equation}\label{eq:VSET}
V_\mathrm{SET} = h\sqrt{\frac{\delta \mu}{\sigma _f\tau_T}}.
\end{equation}
where $\delta \mu$ is the difference in chemical potentials (per volume) between the insulating phase and the conducting phase of the filament, and $\tau_T$ is the thermalization time. In the approximation of 1D heat transfer, one gets $\tau_T = h^2/\chi$. The thermal diffusivity, $\chi = \kappa/ C$ where $C$ is the volumetric heat capacity. Hence $V_{\textrm{SET}}$ can be written as,
\begin{equation}\label{eq:VSET2}
V_\mathrm{SET} \approx \sqrt{\frac{\kappa \delta \mu}{\sigma C}}.
\end{equation}
Using the Einstein model, $C = 3nk_B$ where $n$ is the number density ($N$/$Vol$) of atoms. The chemical potential can be expressed in terms of chemical potential per atom ($\delta \mu_{a}$) as $\delta \mu = n\delta \mu_{a} $. Estimating $\delta\mu _a\approx k_BT$  where $T$ is the temperature at which the filament phase is formed, and adopting the notations of
Widemann-Franz-Lorenz law the SET voltage is then expressed as,
\begin{equation}\label{eq:VSETfin}
V_\mathrm{SET} \approx \sqrt{\frac{L}{3}}T.
\end{equation}

For numerical estimate, we take the above discussed $T = 600$K and $L = 6.67 \cdot 10^{-7}$ W$\Omega$K$^{-2}$ which yields $V_{\textrm{SET}} \sim 0.3$ V which is in the ballpark of measured SET voltages \cite{Fantini 2012, wouters2012}. Perhaps more importantly, Eq. (\ref{eq:VSETfin}) predicts that RRAM devices in which the filament materials have low Lorenz numbers can operate at the correspondingly lower SET voltages.

\section{Conclusion}
We have presented the accurate COMSOL modeling of thermal transport in filamentary devices in comparison with simplified analytical models, one neglecting the radial heat propagation, while the other approximating the temperature distribution as uniform in transversal direction. Our results show that for the typical device parameters, both approximations somewhat overestimate the filament temperature, although the errors ($\lesssim$ 10\%) are not very significant. In particular, we have shown that the model of 1D (along the filament) thermal transport is surprisingly accurate when amplified with a realistic heat transport through the electrode suggested here. These simplified models can serve as a convenient express analysis tool. Our predicted values of filament temperature fall in the ballpark of the earlier measured and modeled values \cite{kim2013,yalon2014,yalon2015}; some differences can be attributed to the particular structural and material parameter choices, and the effects of interfacial resistances \cite{kencke2007} neglected here.

Also, we have demonstrated that the Wiedemann-Franz-Lorenz law can be used in device analysis when the value of Lorenz parameter is taken to correspond the experimental data on electric and thermal conductivity, which can result in significant deviations from its Sommerfeld's value.

Finally, we have presented an approximate expression for the SET voltage in filamentary RRAM structures, possessing high degree of universality and correctly predicting the measured values. That expression points at the filament materials wilt low Lorenz numbers as candidates for the future low switching voltage devices.

\section{Acknowledgement}
We are grateful to I. V. Karpov, R. Kotlyar, and V. I. Kozub for useful discussions.


\appendices
\section{COMSOL Model}\label{appendix0}
Our COMSOL algorithm is as follows.\\
1)  Open the \textbf{Model Wizard}.\\
2) Choose \textbf{2D Axisymmetric} as \textbf{Space Dimension}\\
3) Choose 2D \textbf{AC/DC} module and add \textbf{Electric Currents} sub module in \textbf{Physics}.\\
4) Choose \textbf{Heat Transfer Module} and add \textbf{Heat Transfer in Solids} submodule in \textbf{Physics}.\\
5) Create the \textbf{Geometry} of the MIM structure as in Fig(\ref{Fig:MIM}).\\
6) Create \textbf{Blank Materials} in the \textbf{Materials} node and add material parameters given in Table I to create the required materials.\\
7) Assign the materials to the corresponding domains.\\
8) The \textbf{Heat Transfer in Solids} submodule will have four different necessary default subnodes - add \textbf{Temperature} boundary condition and select the top boundary of the top electrode and bottom boundary of the bottom electrode  and choose 300K in the user defined temperature section.\\
9) The \textbf{Electric Currents} submodule also has four different necessary default subnodes - add \textbf{Electric Potential} boundary condition, select the top boundary of the top electrode and set 0.5 V in the \textbf{Electric Potential box}, add \textbf{Ground} boundary condition, select the bottom boundary of the bottom electrode.\\
10) In \textbf{Multiphysics} node select all the domain in \textbf{Electromagnetic Heat Source} sub-node to couple the \textbf{Electric Currents} submodule and \textbf{Heat Transfer in Solids} submodule,\\
11) Create \textbf{Mesh}.\\
12) Select \textbf{Study}.\\
13) Obtain results in desired form from the \textbf{Results} node.

\section{1D Heat Transfer}\label{appendix1}
In this appendix, we present the solution to the 1D heat transfer problem discussed in Section \ref{sec:1D}.
Multiplying Eq.(\ref{eq:heateq2}) with $dT/dz$ and integrating yields,
\begin{equation}\label{appen:Integrate1}
\int_{0}^{(T'(z))^{2}}d\left(\frac{dT}{dz}\right)^{2}=-\frac{2E^2}{L}\int_{T_\mathrm{max}}^{T}\frac{dT}{T}.
\end{equation}
or,
\begin{equation}\label{appen:Integrate2}
\frac{dT}{dz} = \pm\sqrt{\frac{2E^{2}}{L}}\sqrt{\ln\left(\frac{T_\mathrm{max}}{T}\right)}
\end{equation}
where the origin is at the center of filament.
Define $t\equiv T/ T_\textrm{max}$. Integrating Eq. \ref{appen:Integrate2} from $t(0)= 1$ to $t(z) = T/T_\textrm{max}$ gives the temperature distribution in the filament.
\begin{equation}\label{appen:Integrate4}
\int_{1}^{T/T_\mathrm{max}}\frac{dt}{\sqrt{\ln(1/t)}}=\pm\frac{1}{T_\mathrm{max}}.\sqrt{\frac{2E^{2}}{L}}\int_{0}^{z}dz.
\end{equation}
With a substitution of $y^2 = \ln(1/t)$, the integration changes to,
\begin{equation}\label{appen:Integrate5}
-2\int_{0}^{\sqrt{\ln(T_\mathrm{max}/T)}}\exp(-y^2)dy = \pm \frac{1}{T_\mathrm{max}}.\sqrt{\frac{2E^{2}}{L}} z.
\end{equation}
The LHS of the Eq. (\ref{appen:Integrate5}) is an error function. Hence we get the following temperature distribution after the integration.
\begin{equation}\label{appen:TempDist1}
T(z)=\frac{T_\mathrm{max}}{\exp\left(\left(\mathrm{erf}^{-1}\left(\frac{\mp z}{T_\mathrm{max}}\sqrt{\frac{2E^{2}}{\pi L}}\right)\right)^2\right)}.
\end{equation}
The maximum temperature can be calculated by integrating Eq. (\ref{appen:Integrate2}) from $t(0)=1$ to $t(h/2) =T_j/T_\textrm{max}$. The integration results in a transcendental equation of the form,
\begin{equation}\label{appen:T_max}
\frac{T_\mathrm{max}}{T_j}\mathrm{erf}\left(\sqrt{\ln\left(\frac{T_\mathrm{max}}{T_j}\right)}\right) = \mp\frac{h}{2 T_j}\sqrt{\frac{2E^{2}}{\pi L}}.
\end{equation}
which can be solved numerically.

\section{Radial Heat Transfer}\label{appendix2}
In this appendix, we present more in detail derivation of Eq. (\ref{eq:divT}) for the radial heat transfer and its solution in Section \ref{sec:rad}.
Writing the heat equation in cylindrical coordinate,
\begin{equation}\label{appen:heat2}
-\kappa_i\left[\frac{1}{\rho}\frac{\partial}{\partial \rho}\rho\frac{\partial}{\partial \rho}T(\rho,z) +\frac{\partial^2}{\partial z^2}T(\rho ,z)\right] = 0.
\end{equation}

Since, $h/H \ll 1$ the ratio $\delta T /T \ll 1$ along $z$-direction, hence replacing $T(\rho,z)$ by average temperature $\overline{T(\rho)}$ and integrating along z-axis from $z=-h/2$ to $z=h/2$, Eq.(\ref{appen:heat2})
\begin{equation}\label{appen:heat4}
\frac{h}{\rho}\frac{d}{d \rho}\rho\frac{d \overline{T(\rho)}}{d \rho} + \left[\frac{\partial T(\rho,z)}{\partial z}|_{z=h/2} - \frac{\partial T(\rho,z)}{\partial z}|_{z=-h/2}\right]=0.
\end{equation}
The second term of the LHS of the Eq.(\ref{appen:heat4}) can be reduced to averages after applying the Neumann boundary condition at the metal-insulator interface. The condition states that,
\begin{equation}\label{appen:BC1}
-\kappa_i\frac{\partial T(\rho,z)}{\partial z}=-\kappa_m\frac{\partial T(\rho,z)}{\partial z}.
\end{equation}
The temperature gradient in the electrode is replaced by the slope $\Delta T /\Delta z$ and also by using the average temperature $\overline{T}(\rho)$ as the boundary temperature of the metal-insulator junction, Eq.(\ref{appen:heat4}) becomes,
\begin{equation}\label{appen:heat5}
\frac{h}{\rho}\frac{d}{d \rho}\rho\frac{d \overline{T}(\rho)}{d \rho} + 2\frac{\kappa_m}{\kappa_i}\left(\frac{T_0 -\overline{T}(\rho)}{H}\right)=0.
\end{equation}
Eq.(\ref{appen:heat5}) can be rewritten as,
\begin{equation}\label{appen:bessel}
\rho^2\frac{d^2\overline{T(\rho)}}{d\rho^2}+\rho\frac{d\overline{T}(\rho)}{d\rho}-\beta^2 \rho^2( \overline{T}(\rho)-T_0) = 0 .
\end{equation}

The solution to Eq. (\ref{appen:bessel}) is the linear combination of zeroth order modified bessel functions $I_0$ and $K_0$,

\begin{equation}\label{appen:tempdist}
\overline{T}(\rho)=C_1 I_0(\beta\rho)+C_2 K_0[\beta\rho]+ T_0.
\end{equation}
where, the constants $C_1$ and $C_2$ is determined by the Neumann boundary condition. We omit $I_0$($C_1 = 0$) from the solution as it increases with $\rho$. Hence the radial temperature distribution is given by,
\begin{equation}\label{appen:tempdist1}
\overline{T(\rho)}=C_2 K_0(\beta\rho)+ T_0.
\end{equation}
The Neumann boundary condition becomes,
\begin{equation}\label{appen:BC1}
-\kappa_f\frac{\partial T(\rho,z)}{\partial z}=-\kappa_i\frac{\partial T(\rho,z)}{\partial z}.
\end{equation}
The filament receives power $P=V^2/R$ from the external bias source, where the electrical resistance $R$ is given by sum of electrodes and filament resistances $R=h/\sigma_f \pi r_f^2+2H/\sigma_m \pi r_f^2$, assuming the current takes the shortest path through electrode to filament having the same cross section area as the filament. The power generated by the filament is given by the product of heat flux ($-\kappa_f dT/d\rho$) and the surface area of the cylinder at $r=d/2$. Using the boundary condition and equating the power received and generated we obtain,
\begin{eqnarray}\label{appen:BC2}
&&C_2=\frac{-1}{\beta K_1(\beta\rho)}\frac{d\overline{T}(\rho)}{d\rho} \qquad\mathrm{and,}\\
&&\frac{d \overline{T}(\rho)}{d\rho}|_{\rho=d/2}=\frac{-V^2 d}{4h\kappa_i(h/\sigma_f+2H/\sigma_m)}.
\end{eqnarray}
Substituting the latter equations, the radial temperature distribution in the insulator becomes,
\begin{equation}\label{appen:tempdist}
\overline{T}(\rho) = \frac{V^2 d}{4 h \kappa_i \beta (h/\sigma_f+2H/\sigma_m)K_1(\beta d/2)} K_0(\beta\rho) + T_0.
\end{equation}

\end{document}